\documentstyle[11pt,aussois2003,twoside,epsf]{article}

\def\edcomment#1{\iffalse\marginpar{\raggedright\sl#1\/}\else\relax\fi}
\reversemarginpar

\begin{document}

\title{Spatial and Temporal Variations of Light Curves in Gravitationally 
Lensed Sources}

\author{Anatolyi Minakov}
\affil{Institute of Radio Astronomy of Nat. Ac. Sci. of Ukraine, 4, Chervonopraporna Str., 
Kharkov 61002, Ukraine}

\author{Victor Vakulik}
\affil{Institute of Astronomy of Kharkiv Nat. University, 35, Sumska Str., 61022 Kharkiv,
Ukraine}

\label{page:first}

\begin{abstract}
The effects of macro- and microlensing on the spatial and temporal characteristics
of images of remote sources, observed through the inner regions of lensing galaxies 
are discussed. A particular attention was given to the case, when microlenses, -- stars,
star-like or planetary bodies, -- are situated  near the critical curves of macrolenses,
- galaxies, stellar clusters , etc. The investigation is of interest for the gravitational
lens (GL) systems, where the lensed images are observed close to the critical curve of a
macrolens. Annular, arched or confluent images should be regarded as an indication to
such a proximity. Numerical simulation allowed to determine structures of critical curves
and caustics, formed by macro and microlenses, and to evaluate possible distortions, caused
by microlenses for various locations with respect to the critical curve of a regular lens.
The difference of our results from those obtained earlier with the standard (linearized)
approach to describe the regular gravitational lens was shown to be the larger the closer
to the critical curve the microlenses are situated. In addition to spatial redistributions
in the visible images, complicated deformation of their light curves occurs in gravitational
lensing of variable in time and extended sources. The magnitudes of the temporal variations
depend not only on the GL parameters (e.g., mass distribution), but on the parameters
of the source as well, such as linear dimension of the emitting region, its location
with respect to the critical curve, and the impulse duration. The analysis shows, that
in this case GL acts as a filter, which passes slow temporal variations without degradations, 
and smoothes the rapid ones.
\end{abstract}

\section{Introduction}

Uncorrelated variations of brightness in the images of a distant source, splitted by the
gravitational field of the lensing galaxy are indicative of microlensing events. The 
phenomenon is clearly seen, e.g., in Q2237+0305, which is known to consist of four 
lensed quasar images within a 1-arcsec radius circle around the center of the lensing 
galaxy. The uncorrelated brightness variations of these images indicate that the effects
of microlenses are mainly observed, but not the intrinsic quasar brightness variations.
Because of significant difficulties of theoretic analysis, microlensing phenomena are
investigated mostly with numerical methods. A standard procedure for numerical simulation
consists in the following. The effect of all microlenses, randomly distributed inside
the galaxy, is described in a twofold manner. The microlenses, which are the most close
ones to the observed image of the source, are specified with
their actual deflection angles. The number of such microlenses is not expected to be larger
than several hundreds, (Paczynski 1986; Seitz \& Schneider 1994; Blandford \& Narayan 1992).
The rest of microlenses, which are far from the line of sight, as well as the smoothly
distributed matter, are specified by introducing a deflection angle, averaged over the
interstellar space. The further simplification of the problem is due to linearization of
the mean deflection angle of the macrolens near a selected lensed image, (Seitz \& Schneider 
1994; Blandford \& Narayan 1992; Schneider et al. 1992). The procedure of the local simplification
works well for isolated source images, which are essentially far from the macrolens critical
curve. It is of interest, however, to investigate also an overall interaction of macro
and microlenses for the lensing galaxy as a whole. Therewith, a possibility appears to
follow the changes of spatial density of microcaustics, and the velocities of their  
displacement depending on their location with respect to the critical curve of the macrolens.
In these cases, typical spatial and temporal scales of microfocusings substantially differ
from those obtained with the linearized approach. The analysis is of a particular importance
in studying microlensing by a star, which is in its turn surrounded by a large amount
of planetary-mass objects. Such a complex GL will be further considered as a system of
macro-milli-microlenses.

In addition to spatial redistribution of brightness in the apparent images, complex
deformations of their light curves also take place in gravitational lensing of sources,
which are extended in space and variable in time. The amount of the deformation depends 
not only on the parameters of the GL, -- mass distribution in the lens, -- but on the 
parameters of the source as well, such as the emitting region dimension, its location 
with respect to the critical curve, and on duration of the source brightness spikes. 
The amount of light curve deformations turned out to be different for different lensed 
images. GL can be regarded as a multi-channel filter, which passes slow brightness variations and
smoothes rapid ones.

The aim of this work was to analyze theoretically and simulate numerically the structures
of critical curves and caustics of complex gravitational lenses, which consist of macro-
and microlenses, for various locations of microlenses with respect to the macrolens critical
curve. Also, peculiarities of gravitational focusing of the source with the arbitrarily 
distributed masses were investigated for the case, when the source is extended in space
and variable in time. To obtain numerical estimates, the results were applied to the Einstein 
Cross Q2237+0305.

\section{Spatial changes}

A complex GL can be presented as a sum of two constituents, with one of them being due
to the masses, contained in compact objects, such as stars, star-like and planet-like
bodies, and the second one, connected with a diffuse matter, such as dusty and dusty-gazeous
clouds. Compact objects, referred to as microlenses (or millilenses) hereafter, can
be defined as point masses $M_i$. A total amount of microlenses $n$ in the lensing galaxy
is sufficiently large. For example, $N\approx 10^{11}\div 10^{12}$ for a spiral galaxy similar
to Q2237+0305. The effect of microlensing was at first analyzed for a single microlens,
and the results were then generalized to the case of many microlenses. Let us introduce
a Cartesian coordinate system $XYZ$ with the origin in the galaxy center of masses, and
the $Z$ axis crossing the point of observations, (Fig. 1). Let $Z_p$ and $Z_s$ be the distances
between the GL, the observer $P$ and the source $S$, respectively. Since $Z_p$ and
$Z_s$ are much larger than the size of the region, where the light beams undergo deflection,
we consider the GL as a thin phase screen, (Bliokh \& Minakov, 1989), which is superposed
with the $Z=0$ plane. Having selected a single microlens from a totality of $N$ microlenses,
located at the lens plane and characterized by the impact parameter $\vec \psi = \vec
\psi_m$, we have for the deflection angle at the lens plane, ($Z=0$):

\begin{equation}
\vec \Theta (\vec \psi)=-{{2r_{gm}}\over {Z_p}}{{\vec \psi -\vec \psi _m}\over{(\vec \psi -
\vec \psi _m)^2}}-\sum ^{N-1}_{i=1}{{2r_{gi}}\over {Z_p}}{{\vec \psi -\vec \psi _i}\over
{(\vec \psi -\vec \psi _i)^2}}+\vec \Theta _{dif}(\vec \psi).
\end{equation} 

Here, $\vec \psi =(\psi _x, \psi_y)$ is the angle of observation, $\vec \psi _i$ a coordinate
and $r_{gi}=2GM_i/c^2$ a gravitational radius of the i-th microlens, $G$ is the gravitational 
constant, and $c$ is the velocity of light. $\vec \Theta _{dif}(\vec \psi)$ is a component 
vector of the deflection angle, connected with the diffusely distributed matter. Supposing 
a multidimensional distribution function of random quantities $r_{gi}, \vec \psi _i$ and a 
distribution law for the diffuse matter in the galaxy to be known, one can obtain the average 
defection angle $<\vec \Theta (\psi)>$ and the variance $<(\delta \vec \Theta)^2 >$ 
from Eq. 1, (see e.g., Bliokh \& Minakov 1989, Schneider et al. 1992).

The GL with the average deflection angle $<\vec \Theta (\psi)>$ is called a regular lens,
or a macrolens. The approximation suits a GL, which has the densities of stellar masses and 
diffuse matter "smeared" over sufficiently large regions of the interstellar space.

Consider a "small" region of space, located near the $m-th$ microlens, which has the size,
from one hand, small enough as compared to the typical interstellar distance $\Psi _{st}$ 
in the particular region, and from the other hand, is larger than or comparable to the 
characteristic angular size of the Einstein ring $\Psi _m= \sqrt {2r_{gm}Z_s/Z_p(Z_p+Z_s)}$
for the given microlens, $(\Psi_m<|\vec\psi - \vec\psi_m|<\Psi_{st}).$ Accounting for
 the smallness of the RMS deflection angle (Bliokh \& Minakov 1989, Schneider et al. 1992),
 one can easily show, that in fact, the second and the third addends in Eg.(1) do not differ
 from the mean deflection angle $<\vec\Theta(\vec\psi)>.$
 
 With all the above noted taken into account, the equation of the complex lens can be
 written as:
\begin{equation}
\vec\psi_s=\vec\psi+{\tilde Z \over {Z_p}}<\vec\Theta(\vec\psi)> - 
\Psi^2_m{{\vec\psi-\vec\psi_m}\over{(\vec\psi-\vec\psi_m)^2}}=\vec F(\vec\psi),
\end{equation}
where $\vec\psi_s$ is the actual location of the "point" source, and $\tilde Z=Z_pZ_s/(Z_p+Z_s).$

The lens equation can be better written in normalized quantities 
$\vec\nu=\vec\psi/\Psi_l,\vec\nu_s=\vec\psi_s/\Psi_l,
\vec\nu_m=\vec\psi_m/\Psi_L.$  It is convenient to determine the normalizing angle 
$\Psi_L$ as the angular radius of the Einstein ring, which would be observed for the given
macrolens, provided it did not contain an asymmetric constituent of the mass distribution.
In the normalized quantities, the equation of lens (2) can be presented as
\begin{equation}
\vec\nu_s=\vec F_M(\vec\nu)-\Gamma_m{{\vec\nu-\vec\nu_m}\over {(\vec\nu-\vec\nu_m)^2}}=
\vec F(\vec\nu),
\end{equation}
where $\vec F_M(\vec\nu) = \vec\nu+{\tilde Z \over Z_p}<\vec\Theta(\vec\nu)>$ is a constituent, 
connected with the macrolens galaxy, and $\Gamma_m=\Psi^2_m/\Psi^2_l<<1$ is a small parameter,
that is approximately equal to the ratio of the microlens mass to the total mass of the
galaxy.
The gravitational lens system Q2237+0305 was selected as an example, for which various
models of mass distribution are known, (see, e.g., Kent\& Falco 1988, Minakov \& Shalyapin
1991, Wambsganss \& Webster 1997). Basing on the observational data, let us choose the
model to be simple enough, but in such a way, that it would be able to produce the observed
cross-shaped structure. Let the regular constituent of the deflection angle $<\vec\Theta(\vec\nu)>$
be presented  as a sum of two terms, with one of them, $\vec\Theta_d(\vec\nu)$ being connected 
with a massive disc, and the other one, $\vec\Theta_c(\vec\nu)$, -- with a compact nucleus.
To simplify the analysis, suppose the nucleus and the disc are spherically and elliptically 
symmetric, respectively, with their centers coinciding. Superpose the directions of $OX$ 
and $OY$ axes with the axes of symmetry of the disc component. Since the typical scale
for changing the surface density of the disc component in Q2237+0305 exceeds noticeably the
region, where four quasar components are observed, - about 2\arcsec, - one may use the following
approximation for $\vec\Theta_d(\vec\nu)$ near the center of the galaxy: $\vec\Theta_d(\vec\nu)
\approx \vec\Theta_d(0)+(\vec\nu \nabla)\vec\Theta_d(0),$ where $\vec\nabla={\partial\over
{\partial v_x}}\vec e_x +{\partial\over{\partial\nu_y}}\vec e_y$, $\vec e_x$ and $\vec e_y$ 
are unit vectors of the Cartesian coordinate system. Having supposed characteristic
size of the galaxy compact nucleus to be smaller than the region, where the four images
are observed, let us approximate the spherical component  by a point mass, equal to the
mass of the galaxy nucleus $M_C$. With the accepted simplifications and suggestions, and
having in mind, that for the chosen model of the disc $\Theta_d(0)=0,$ the regular component
of the macrolens equation can be brought to a form, (Minakov \& Shalyapin 1991):
\begin{equation}
\vec F_M(\vec\nu)=\vec\nu(1-{1\over {\nu^2}})+\alpha\nu_x\vec e_y-\alpha\nu_y\vec e_y.
\end{equation}

Here, $\alpha$ is a dimensionless parameter that determines the amount of asymmetry, 
introduced by the disc component near the galaxy center. For the Q2237+0305 model, the
normalizing angle $\Psi_l\approx 0.9\arcsec,$ $\Gamma_m$ can be estimated to be $\Gamma_m\approx 
10^{-9}\div 10^{-10}$, and the asymmetry parameter can be selected equal to $\alpha\approx 0.16.$
Therefore, the inequality $\Gamma<<\alpha<<1$ holds.

The equation of the critical curve of a GL $\vec\nu=\vec\nu_{cr}$ is obtained from the
condition that Jacobian of transformation from variables $\vec\nu_s$ to variables $\vec\nu$
is equal to zero:
\begin{equation}
q^{-1}(\vec\nu)=|{{\partial(F_x,F_y)}\over \partial{(\nu_x,\nu_y)}}|=0, 
\end{equation}
where $q(\vec\nu)$ is the amplification factor.

If we substitute the solution of $\vec\nu_{cr}$, found from (5), to the initial lens equation
(3), we will obtain the equation of caustic in the source plane:
\begin{equation}
\vec\nu_{cs}=\vec F_M(\vec\nu_{cr})-\Gamma_m{{\vec\nu_{cr}-\vec\nu_m}\over{(\vec\nu_{cr}-
\vec\nu_m)^2}}.
\end{equation}

As the first step of the further analysis, let us determine the critical curve 
$\vec\nu=\vec\nu^M_{cr}$ and caustic $\vec\nu_{cs}=\vec\nu^M_{cs}$ of the macrolens,
having supposed $\Gamma_m=0$ in Eq. (3). For example, for the model (4) in the polar 
coordinates $\vec\nu=(\nu,\varphi,)$ in a linear approximation with respect to a small 
parameter $\alpha$, we have:
\begin{equation}
\nu^M_{cr}(\varphi)\approx 1+{\alpha\over 2}\cos 2\varphi,
\end{equation}
with the azimuth angle counted off the $X$ axis.

The equation of caustic of the regular GL in the source plane in the same approximation
is:
\begin{equation}
\nu^M_{cs x}(\varphi)=2\alpha \cos^3\varphi, \qquad \nu^M_{cs y}(\varphi)
=-2\alpha\sin^3\varphi.
\end{equation}

The model of a regular GL under consideration is seen to produce an oval critical curve 
in the lens plane, with an approximately unit average radius, - $\nu^M_{cr}\approx 1,$
while the corresponding caustic curve in the source plane is diamond-shaped, with the
typical size of $|\nu^M_{cs}|\approx 2\alpha.$ It can be easily shown, that four images 
will emerge in projection of the point source $S$ inside the diamond caustics (8), 
with two images outside the critical curve, and two ones inside the curve (7) in the lens
plane, (see, e.g. Minakov \& Shalyapin 1991). Because of smallness of $\alpha$, the
images are located close to the critical curve (7).

\section{Effect of microlenses}

Regarding the equations of the macrolens critical curve and caustic to be known, one
may investigate deformations of the curves for various locations of a microlens $\vec\nu_m$
with a specified parameter $\Gamma_m$. A simple method was applied to numerically solve
Eq.(5). The problem of finding the roots of equation $q^{-1}(\vec\nu)=0$ can be substituted
by a more simple one. Thus, it is easy to numerically determine the regions of positive, 
$q^{-1}(\vec\nu)>0,$ and negative,  $q^{-1}(\vec\nu)<0,$ values of $q^{-1}(\vec\nu).$ 
Excluding singularities, connected with the point mass approximation, the problem is 
reduced to determining the boundary between the positive and negative values of 
$q^{-1}(\vec\nu).$ The simulation has shown, that a behavior of critical curves and 
the relevant caustics of a complex GL depends not only on the distance between a microlens 
and the critical curve of a regular GL, but also on the sector, where the microlens moves. 
Critical curves and caustics for the Q2237+0305 model were calculated for different 
locations of the microlens, - inside and outside the critical curve of the regular GL, 
and exactly in it, -- and are shown in Fig. 3a,b,c. The cases of the microlens
displacement along the $X$ and $Y$ axes were examined. To ensure better view of macro-
and microcurves behavior in the same picture, a sufficiently large value for $\Gamma$
was selected, $\Gamma=0.001.$ The curves for $\Gamma=10^{-9}$ will be also presented below.

For a microlens outside the critical curve of a regular GL, -- Fig.2a, $q^{-1}(\vec\nu)>0$ 
region, -- the curve $\vec\nu=\vec\nu_{cr}$ consists of two closed curves. One of them
is a slightly deformed macrolens critical curve $\vec\nu^M_{cr}$, and the other one appears
around the microlens. The similar behavior of the critical curve of a complex GL is observed
also when the microlens is inside the critical curve of a regular macrolens, -- Fig. 
2b, $q^{-1}(\vec\nu)<0.$ In contrast to the case 2a, however, the microlens critical curve
is not a single deformed oval around the microlens (Einstein "ring"), but consists of
two closed curves on either side of the microlens. The most interesting case, when the
microlens is situated right at the macrolens critical curve, is shown in Fig.2c. The critical
curves of macro- and microlenses merge into a joint continuous curve in this case.

A numerical analysis allows to clearly demonstrate a complex deformation of critical
and caustic curves, but mathematic analysis is needed to obtain qantitative estimates of 
the expected effects, that would make it possible to follow their dependence on the model
parameters. Determining of the real roots of Eq.(5) for the given values of displacement
$\vec\nu_m$ and parameter $\Gamma_m<<1$ is the major task of the analytic investigation.

The following conclusions have been made in the present theoretic analysis, applied to
a particular case of Q2237+0305, (Minakov \& Vakulik 2000):

\begin{itemize}
\item A microlens distorts a critical curve $\vec\nu^M_{cr}$ and caustic $\vec\nu^M_{cs}$
of a regular GL. The amount of deformation is proportional to $\Gamma_m$
and decreases rapidly as the distance between the microlens and the macrolens critical
curve is growing, see Fig. 3. Since rather large number of microlenses may be present 
near the critical curve simultaneously, their joint effect will cause "blurring" of 
boundaries of the critical curve and caustic of the macrolens. In Fig. 4, the structure of 
critical curves and caustics for Q2237+0305 is shown for the case of three hundred 
microlenses, located near the $\vec\nu^M_{cr}$.

\item A structure and behavior of critical curves near the microlens depends not only on the
distance to the critical curve of the regular GL, but on the sector of angles as well.
Characteristic dimensions of critical curves and caustics are proportional to $\sqrt\Gamma_m.$

\item The strongest deformation of critical curves and caustics, proportional to $\Gamma^{1/3}_m$
is observed when microlenses are situated very close to the critical curve of the regular
GL. The curves for macro- and microlenses merge together to form complex continuous
curves.

\item The difference between the results, obtained with the proposed approach and those of
the linear approximation is the larger the closer a microlens to the critical curve, as
can be seen from Fig. 5. 

\item The effect of interacting microlenses increases noticeably in their approaching
the critical curve of a regular GL. 

\item If even microlenses are distributed homogeneously, spatial distribution of their
microcaustics is substantially inhomogeneous near the macrocaustic, see Fig. 4. Such
an inhomogeneity results in spatial scale of microlensing being dependent on the macroimage
location with respect to the macrolens critical curve.
\end{itemize}

The theoretic analysis and the obtained estimates are capable of examining more complicated
situations as well. Suppose, for example, that a complex GL system consists of a macrolens 
galaxy, millilens star, with the mass of order of $M_{\sun}$, and of a large number of 
microlenses, -- planetary bodies with the masses of $10^{-5}M_{\sun}$. If a millilens is 
situated far from the critical curve of the macrolens galaxy, the effect of focusing can 
be more easily analyzed in the coordinate system superposed with the millilens. In doing so, 
the gravitational field of the galaxy may be considered in the linear approximation, while 
the effect of microlenses, randomly distributed around the millilens, should be considered 
in the way similar to that described above. It is only necessary to renormalize the angles 
to the quantity $\Psi_l$, corresponding to the Einstein ring of the millilens. The parameter 
$\Gamma_{\mu}$ will be proportional to the ratio of the mass of the $\mu$-th microlens to 
the mass of the millilens.

\section{Temporal changes}
 
 A complicated redistribution of brightness in the observed image takes place in gravitational
 focusing. If parameters of the source and the lens do not change in time, the brightness
 distribution $I_p(\vec\psi),$ observed through the GL is related to the actual source brightness
 distribution $I_s(\vec\psi_s)$ by a convolution-like formula, (Bliokh \& Minakov 1989):
\begin{equation}
I_p(\vec\psi)=\int^{\infty}_{-\infty}I_s(\vec\psi_s)\delta\left [\vec\psi_s-\vec F(\vec\psi)
\right]
d\vec\psi_s.
\end{equation}
Here, $\delta\left[\vec\psi\right]$ is the delta-function. The formula represents 
mathematically the property inherent in all the nonabsorbing lenses to remain the 
brightness along a particular light ray unchanged.

Several compact bright images (macroimages), separated by a few arcseconds are usually 
observed in actual gravitational lenses. The macroimages  have dimensions approximately
coinsiding with those of the source emitting regions. For example, they are about $10^{-4}
\div 10^{-6}$ arcseconds for quasars in optical wavelegths. For radio wavelengths, angular
dimensions of macroimages may be several orders of magnitude larger. Contemporary instruments,
-- telescopes, radio telescopes, -- do not have a resolution high enough to examine the
fine structure of macroimages $I_p(\vec\psi)$. Therefore, the source  brightness, integrated 
over angles will be observed:
\begin{equation}
J_p=\int^{\infty}_{-\infty} I_p(\vec\psi)d\vec\psi.
\end{equation}
Amplification of the source brightness, resulted from gravitational focusing of a stationary
source, is characterized by the amplification factor $q$, which is defined as:
\begin{equation}
q=J_p/J^{(0)}_p,
\end{equation}
where $J_p^{(0)}$ is the source brightness, which would be observed without the GL. If
$J_p$ is ment to be a total brightness of the whole distorted image of the source, then
$q$ determines the total amplification factor of the lens. If $J_p$ is the
brightness of the $i-$the macroimage, then $q$ means the amplification factor of the $i-$th
macroimage, $q_i$. In this case, coefficients $q$ and $q_i$ are associated by a simple
relationship:
\begin{equation}
q=\sum _iq_i,
\end{equation}

In the more general case, when the brightness of the source varies not only in angles,
but also in time, -- $I_s=I_s(\vec\psi_s, t)$, -- while the GL parameters are virtually
unchanged, equations (9) and (10) can be generalized to take the form (Minakov \& Nechaev
1998):
\begin{equation}
I_p(\vec\psi, t)=\int^{\infty}_{-\infty}I_s\left[\vec\psi_s,t-t_p(\vec\psi, \vec\psi_s)\right]
\delta\left[\vec\psi_s-\vec F(\vec\psi)\right]d\vec\psi,
\end{equation}
\begin{equation}
J_p(t)=\int^{\infty}_{-\infty}I_p(\vec\psi,t)d\vec\psi.
\end{equation}
$t_p(\vec\psi,\vec\psi_s)$ is a signal propagation time along a virtual ray, that
connects an arbitrary point of the source $(\vec\psi_s)$ with an arbitrary point at the
lens plane $(\vec\psi)$, and, further, with the observer $P$. Taking into account that
the local velocity of the signal propagation in the weak gravitational field with a potential
$\Phi$, -- $| \Phi| /c_2<<1,$ -- is equal to $c/n(\vec r)$, and introducing the effective 
refraction index of the "medium" $n(\vec r)$, (Bliokh \& Minakov 1989), the propagation time 
$t_p(\vec\psi,\vec\psi_s)$ can be determined by integrating $n(\vec r)$ along a curvilinear
ray $L$, that associates arbitrary points of the source, the lens and the observer: 
\begin{equation}
t_p=c^{-1}\int_L n(\vec r)dl.
\end{equation}
The effective refraction index can be representd as $n(\vec r)=1-2\Phi(\vec r)/c^2.$

Using  a "thin" lens and paraxial optics approximation, intergration along the
curvilinear ray in (15) can be replaced by integration along the lines, connecting  
a point element at the source surface with a point element at the lens plane, 
and the lens point with the observer. After simple transformations, we have the 
following expression:
\begin{equation}
t_p(\vec\psi, \vec\psi_s)=c^{-1}\left[Z_p+Z_s+{{Z_p(Z_p+Z_s)}\over Z_s}
\left({1\over 2}{{Z_p+Z_s}\over{Z_p}}\psi^2_s+{1\over2} \psi^2 -\vec\psi \vec\psi_s\right)\right]+
t_{gr}(\vec\psi).
\end{equation}

The first addend in this formula is the propagation time for the signal in the empty space
along the broken path "source-lens-observer", see Fig. 1. The second term, $t_{gr}$,
is due to the presence of gravitational field. For the weak gravitational field, 
$t_{gr}(\vec\psi)$ can be calculated by integrating of a gravitational addition $2\Phi/c^2$
in (15) along the rectilinear rays. It is easy to show, that $t_{gr}(\vec\psi)$ can be 
reduced to a form:
\begin{equation}
t_{gr}(\vec\psi)={{2G}\over{c^3}}\int_W d\vec\psi\arcmin \sigma(\vec\psi\arcmin)\ln{{4Z_s}
\over{Z_p(\vec\psi\arcmin-\vec\psi)^2}}+const,
\end{equation}
where $\sigma(\vec\psi)$ is the given surface mass
density distribution for the lens, $\left[ \sigma\right] =kg/rad^2$, and integration is 
carried out over the whole surface of the GL in projection to the sky, and $const$ is 
an insignificant addition, affecting only a zero point of the time. Eq. (16) and (17)
hold for the low-redshift GL systems, when the cosmology factor can be neglected. 

Eq. (13) has a simple physical meaning. The point spread function of the lens,
$\delta\left[\vec\psi_s-\vec F(\vec\psi)\right],$ 
associates every particular angle of observation with a certain "point" element
of the source surface, having an actual angle $\vec\psi_s=\vec F(\vec\psi)$ and a surface
brightness $I_s(\vec\psi_s, t).$ Temporal changes of $I_s(\vec\psi_s, t)$ will be recorded
by the observer at an angle $\vec\psi$ with a time delay of \quad $t_p(\vec\psi)=t_p(\vec\psi,
\vec\psi_s=\vec F(\vec\psi))$. The exact coincidence between the temporal variations of 
the source brightness and brightness of its images is seen to occur only for the point 
sources. In focusing extended sources, their actual light curves will differ from the light
curves of the lensed images.

To simplify the further analysis, we studied a variable source with the Gaussian brightness
profile:
\begin{equation}
I_s(\vec\psi_s, t)={{\tilde J_0}\over {2\pi\Psi_0^2}}
\exp \left[-{{(\vec\psi_s-\vec\Psi_s)^2}\over
{2\Psi_0^2}}\right] \cdot f(t,\vec\psi_s).
\end{equation}
where $\Psi_0$ and $\vec\Psi_s$ are angular dimension and angular coordinate of the source 
brightness maximum, respectively, and $f(t,\vec\psi_s)$ is a dimensionless function of the 
order of unity, accounting for the temporal variations of brightness. It is supposed 
further, that the source varies in time as a whole according to:
\begin{equation}
f(t, \vec\psi_s)={t\over {T_s}}\exp \left(-{{t^2}\over {2T_s}}\right)\cdot h(t),
\end{equation}
where $T_s$ is a characteristic duration of the brightness flash, $h(t)=1$ at $t\geq 0$,
and $h(t)=0$ at $t<0.$

This model refers to the source, which flashed at $t=0$, and then is gradually fading,
with a characteristic decay time $T_s.$ The $\tilde J_0$ quantity in Eq. (18) is the full
energy, emitted by the source:
\begin{equation}
\tilde J_0=\int^{\infty}_{-\infty}d\vec\psi_s\int^{\infty}_{-\infty}dt I_s(\vec\psi_s,t).
\end{equation}

For the given model (18), and according to (13), the observed brightness distribution
is described by the expression:

$$I_p(\vec\psi,t)={{\tilde J_0}\over {2\pi\Psi_0^2T_s}} \exp \left[-{{(\vec\Psi_s-\vec F(\vec\psi))^2}
\over {2\Psi_0^2}}\right]\cdot h\left(t-t_p(\vec\psi)\right)\cdot$$
\begin{equation}
\cdot\left(t-t_p(\vec\psi)\right) 
\exp \left[-{1\over {2T_s}}\left(t-t_p(\vec\psi)\right)^2\right],
\end{equation}
where $t_p(\vec\psi)=t_p\left(\vec\psi,\vec\psi_s=\vec F(\vec\psi)\right).$

Normalized quantities $\vec\nu=\vec\psi/\Psi_l,$ $\vec\nu_s=\vec\Psi_s/\Psi_l,$ 
$\nu_0=\Psi_0/\Psi_l,$ $\vec F(\vec\nu)=\vec F(\Psi_l\vec\nu)/\Psi_l$, and  normalized 
time intervals $\tau=t/T_s$ and $\tau_p(\vec\nu)=t_p(\Psi_l\vec\nu,\Psi_l 
\vec F(\vec\nu))/T_s$ are more convenient to analyze gravitational focusing. 
In the normalized coordinates, the source brightness distribution (21) is reduced to a
form:
\begin{equation}
I_p(\vec\nu, \tau)={{\tilde J_0}\over {2\pi\nu^2_0}}h\left(\tau-\tau_p(\vec\nu)\right)
\left(\tau-\tau_p(\vec\nu)\right)
\exp{\left[-{1\over 2{\nu_0^2}}\left(\vec\nu_s-\vec F(\vec\nu)\right)
^2 -{1\over 2}\left(\tau-\tau_p(\vec\nu)\right)^2\right]}.
\end{equation}

We can get the light curve of the observed image $J_p(\tau)$ having integrated Eq. (22)
over all possible angles of observations:
\begin{equation}
J_p(\tau)=\int^{\infty}_{-\infty}I_p(\vec\nu, \tau)d\vec\nu.
\end{equation} 
The full energy of the observed splash of brightness (impulse hereafter) can be determined as
\begin{equation}
\tilde J_p=\int^{\infty}_{-\infty}d\tau)J_p(\tau).
\end{equation} 

\section{Calculation of light curves}

For the GL systems observed in optics, $\nu_0<<1$ holds as a rule, excluding
rather extended radio sources, where $\nu_0$ may be of the order of unity or even
larger. It is known from the theory of GL, that the focusing effect of the gravitational
field is negligible for the sources with $\nu_0>1$, therefore, this case will not be analyzed
further. According to (21), for $\nu_0<<1$, radiation gets into the point of observation 
only from the small regions of space, situated near the point $\vec\nu=\vec\nu_i$, where
the index of power in the exponent becomes zero, i.e. when $\vec\nu_s=\vec F(\vec\nu_i)$. 
The roots $\vec\nu=\vec\nu_i$ of this equation, - lens equation (3), - determine the coordinates
of brightness peaks in the images, formed by the GL. Supposing, that $N$ separate images
of the source are formed by the lens, - $i=1, 2, \dots, N$, - let us introduce small 
deviations $\Delta\vec\nu$ from maxima $\vec\nu_i$, - $\vec\nu=\vec\nu_i+\Delta\nu$, - 
and make the following simplifications, (Minakov et al. 2001):
$$\vec\nu_s-\vec F(\vec\nu_i+\Delta\vec\nu)\approx (\Delta\vec\nu \vec\nabla)\vec F(\vec\nu_i),$$
\begin{equation}
\tau-\tau_p(\vec\nu_i+\Delta\vec\nu)\approx \tau\arcmin_i-(\Delta\vec\nu\vec\nabla)\tau_p(\vec\nu_i),
\end{equation} 
where $\tau\arcmin_i=\tau-\tau_p(\vec\nu_i)$. With simplifications (25) taken into
account, the distribution of brightness (13) near the $i$-th image maximum may be 
presented as
$$I_{pi}(\Delta\vec\nu, \tau\arcmin_i)\approx {{\tilde J_o}\over{2\pi\nu^2_0}}h
(\tau\arcmin_i-\tilde A_i\Delta\nu_x-\tilde B_i\Delta\nu_y)\cdot(\tau\arcmin_i-
\tilde A_i\Delta\nu_x-\tilde B_i\Delta\nu_y)\times$$
$$\times\exp {\left[-{1\over 2}(\tau\arcmin_i-\tilde A_i\Delta\nu_x-\tilde B_i
\Delta\nu_y)^2\right]}\times$$ 
\begin{equation}
\times \exp{\left[-{1\over{2\nu_0^2}}\left((\tilde C_i\Delta\nu_x+\tilde D_i\Delta\nu_y)
\vec e_x+(\tilde D_i\Delta\nu_x+\tilde E_i\Delta\nu_y)\vec e_y\right)^2\right]},
\end{equation} 
where
\begin{equation}
\tilde A_i={{\partial\tau_p}\over {\partial \nu_x}};\enskip \tilde B_i={{\partial\tau_p}
\over{\partial \nu_y}};\enskip \tilde C_i={{\partial F_{x}}\over{\partial \nu_x}}; 
\enskip\tilde D_i={{\partial F_{y}}\over {\partial \nu_y}}; \enskip \tilde E_i=
{{\partial F_{x}}\over{\partial\nu_y}}={{\partial F_{y}}\over{\partial\nu_x}}.
\end{equation} 
All the derivatives in (27) are calculated in the point $\vec\nu=\vec\nu_i$. Having in
mind the introduced simplifications, after simple calculations, one
can get the following asymptotic approximation for the light curve of the $i$-th image:
$$J_{pi}(\tau\arcmin_i)\approx\int^{\infty}_{-\infty}d(\Delta\vec\nu)I_{pi}(\Delta\vec\nu,
\tau\arcmin_i)
={{q_i\tilde J_0}\over{\sqrt{2\pi}}}\cdot{\sqrt{R_i}\over {1+R_i}}\cdot exp{\left(-{R_i
\over 2}\cdot\tau\arcmin^2_i\right)}\times$$
\begin{equation}
\times \left[ 1+\sqrt{\pi}\cdot{{R_i\tau\arcmin_i}\over{\sqrt{2(1+R_i)}}}
\cdot\exp 
\left({{R_i^2\tau\arcmin^2_i}\over{2(1+R_i)}}\right)\cdot
\left(1+\Phi\left({{R_i\tau\arcmin_i}\over{\sqrt{2(1+R_i)}}}\right)\right)\right].
\end{equation} 

The following designations are introduced here:
\begin{equation}
R_i={1\over{\nu^2_0q_i^2H_i}};\enskip H_i=(\tilde C_i^2+\tilde D_i^2)\tilde B_i^2+
(\tilde D_i^2+\tilde E_i^2)\tilde A_i^2-2\tilde A_i\tilde B_i\tilde D_i(\tilde C_i+\tilde E_i);
\end{equation} 
\begin{equation}
q_i=| \tilde C_i\tilde E_i-\tilde D_i^2|^{-1},
\end{equation} 
and $\Phi(x)$ is the probability integral. The introduced quantity $q_i$ can be easily shown
to represent the amplification factor for the $i$-th image of a stationary source.

As it follows from simple physical considerations, amplification of the initial impulse
$\tilde J_0$ in gravitational focusing originates only from the spatial redistribution
of  brightness, determined by the lens amplification factor $q$. Thus, a condition 
$\tilde J_{pi}=q_i\tilde J_0$ must hold. Indeed, having integrated (28) over the whole
region of variation of $\tau\arcmin_i$, $(-\infty<\tau\arcmin_i<\infty)$, we get the full
energy of the focused impulse, arrived to the point of observations from the $i$-th image:
\begin{equation}
\tilde J_{pi}=\int^{\infty}_{-\infty}J_{pi}(\tau\arcmin_i)d\tau\arcmin_i=q_i\tilde J_0.
\end{equation} 

For the values $R_i>>1$, we get the following approximate expression for
$J_{pi}(\tau\arcmin_i)$, having used the asymptotic expression of $\Phi(x)$:
\begin{equation}
J_{pi}(\tau\arcmin_i)=q_i\tilde J_0\cdot\tau\arcmin_i\cdot\exp\left(-{{\tau\arcmin^2_i}\over
2}\right), 
\quad \tau\arcmin_i>0.
\end{equation} 

The light curve of the observed $i$-th image is seen to be formed from the initial undisturbed 
light curve
\begin{equation}
J_p^{(0)}(\tau\arcmin)=\tilde J_0\cdot\tau\arcmin_i\cdot\exp\left(-{{\tau\arcmin^2_i}\over
2}\right), \quad \tau\arcmin_i>0
\end{equation} 
by multiplying the latter by the amplification factor $q_i$. It should be noted however,
that such a simple effect of a GL is observed only for $R_i>>1.$  At small values of $R_i,$
the lens will distort the initial light curve of the source. In addition, the amount of
distortions will be different for different images because of the existing discrepancy
in the values of $R_i.$ Relationships $Q(\tau\arcmin)=\sqrt{2\pi} J_p(\tau\prime)/\tilde J_0q$
for different $R$ are presented in Fig. 7.

We investigated focusing of a quasar radiation by a regular component of the Q2237+0305
system as an example. For this system, $\vec F(\vec\nu)=\vec F_M(\vec\nu)$ and is described
by Eq. (4), while the signal time delay $\tau_p(\vec\nu)$, disregarding an insignificant
additive term, is described by expression:
$$\tau_p(\vec\nu)=\tau_0+\tau_g{{Z_s}\over {2\tilde Z}}
\left[(1+\alpha^2)\nu^2-2+{1\over{\nu^2}}
+2\alpha\left(1-{1\over{\nu^2}}\right)(\nu_x^2-\nu_y^2)\right]-$$
\begin{equation}
+\tau_g\left[-{{\nu^2}\over 2}+1-\ln
\nu -{\alpha\over 2}(\nu_x^2-\nu_y^2)\right],
\end{equation} 
where $\tau_g=2r_g/cT_s,$ \enskip $r_g\sim 3\cdot10^9\div 3\cdot10^{10}km$ is a gravitational
radius of the galaxy nucleus, and $Z_s/\tilde Z\approx 10$ is the relative distance, (Kayser,
Refsdal \& Stabell 1986). It is easy to show, that there are four points in Eq.(34), where
$\tau_p(\vec\nu)$ is minimal. For $\alpha=0.16$ and $Z_s/\tilde Z=10,$ coordinates of
these points are, respectively:
$$\nu_{1,3}=\sqrt{{Z_s}\over {Z_s(1+\alpha)-\tilde Z}}\approx 0.97, \enskip \varphi_1=0,
\enskip \varphi_3=\pi;$$
\begin{equation}
\nu_{2,4}=\sqrt{{Z_s}\over {Z_s(1-\alpha)-\tilde Z}}\approx 1.16, \enskip \varphi_2={\pi\over
2}, \enskip \varphi_4={{3\pi}\over 2}.
\end{equation} 
\begin{table}
\caption{Values of normalized angles and parameters for Q2237+0305 model}
\begin{tabular}{cccccccc}
\tableline
 Image &$\nu_x$&$\nu_y$&$\nu_x(\vec\nu_s=0)$&$\nu_y(\vec\nu=0)$&$q_i$ &$\gamma_i$&$\beta_i$\\
\tableline
    A  & 0.325 &-1.031 & 0                  &-1.08             & 2.261& 0.251   & 0.446 \\
    B  & 0.279 & 1.041 & 0                  & 1.08             & 2.244& 0.258   & 0.443 \\
    C  &-0.88  &-0.044 &-0.94               & 0                & 0.969& 5.21    & 0.774 \\
    D  & 0.994 &-0.099 & 0.94               & 0                & 3.395& 0.015   & 0.526 \\
 Nucleus&  0   & 0     & 0                  & 0                &      &         &       \\
\tableline
\tableline
\end{tabular}
\end{table}

Presense of minima in $\tau_p(\vec\nu)$ indicates, that temporal changes in the source
will be observed in the lensed images in the directions of points (35) at first, and then
in the A, B, C and D images. 

For the obtained $\tau_p(\vec\nu)$, (34), a dimensionless parameter $H_i$ in (29) can be 
represented as $H_i=\tau^2_g\gamma(\vec\nu_i),$ where $\gamma(\vec\nu_i)=\gamma_i=H_i/\tau_g^2$
does not depend on $\tau_g$ and determines only by the value of $\alpha,$ relative distance
$Z_s/\tilde Z$, and by the coordinates of the four images. For the given
coordinates of the images, the corresponding values of parameters $q_i,$ $\gamma_i$
and $\beta_i=[\tau_p(\vec\nu_i)-\tau_0]/\tau_g,$ \enskip($i=A,B,C,D$) were calculated. 
The results are presented in Table 1. The coordinates of the observed lensed images from 
(Crane et al. 1991) are also presented here, as well as the coordinates of the same images 
for the central source $(\vec\nu_s=0).$

Further calculations were made for the angular dimension of the source of
$\nu\approx 10^{-3},$ and for two values of $\tau_g=0.1, \tau_g= 600.$ In Fig. 7, the 
individual light curves $J_{pi}(\tau\arcmin_i)$ (left), and a total one $J_{p\Sigma}=
\sum J_{pi}(\tau\arcmin_{pi})$ for $\tau_g=0.1$, (right) are presented.

It is evident, that there is no possibility to detect separate images in the total light
curve at small values of $\tau_g.$ The components become distinguishable in the total light
curve for sufficiently large values of $\tau_g.$ For example, a total light curve for
$\tau_g=600$ is shown in Fig. 8.  

The discussed light curves referred only to a variable component of the source radiation.
In reality, the radiation can be represented as a sum of a stationary and variable constituent.
In this case, with (31) taken into account, the initial light curve would be given by
the following expression:
\begin{equation}
J^{(0)}_{p\Sigma}(\tau\arcmin)=J_s+\tilde J_0\tau\arcmin\exp{\left(-{1\over2}
\tau\arcmin_i^2\right)},
\end{equation}
where $J_s=const$ is a stationary component of the source radiation. The following total
brightness as a function of time will be recorded in the point of observation:
\begin{equation}
J_{p\Sigma}(\tau\arcmin)=qJ_s+\tilde J_0\sum _iq_i\tau\arcmin_i\exp{\left(-{1\over
2}\tau\arcmin^2_i\right)}=
J_s\left[q+{{\tilde J_0}\over{J_s}}\sum_i q_i\tau\arcmin_i
\exp\left(-{1\over 2}\tau\arcmin^2_i\right)\right],
\end{equation}
where $q=\sum q_i\approx 8.9$ is a total amplification factor of all the four
images of Q2237+0305.

\section{Discussion and conclusions}

Some conclusions can be made from the analysis of gravitational focusing of the extended
source, that varies in time.

If a gravitational lens forms several images of the source, the light curve of an individual
image $J_p(\tau\arcmin)$ will depend on the value of parameter $R$, which accounts simultaneously
for geometry of the task (distances from the observer to the source and to the lens, $Z_p$
and $Z_s$ respectively),  parameters of the lens (a total mass and its distribution),
and the source parameters ($\Psi_0, \vec\Psi_s, T_s$). For the model of GL Q2237+0305
under consideration, -- a lens with a quadrupole, -- the value of $R$ was estimated to be
$R\sim (\nu_0\tau_g)^{-2}.$

If $R>>1,$  ($\nu_0\tau_g<<1$ -- a "small" source, "weak" lens or "long" impulse), the
light curve $J_p(\tau\arcmin)$ does not depend on $R$ any more, and differs from the 
undisturbed light curve $J_p^{(0)}(\tau\arcmin)$ by a multiplier $q$ only.

For small values of $R$, when $R<<1$, ($\nu_0\tau_g>1$ -- "extended" source, "strong" lens,
or "short" impulse), a smoothing action of the lens should be taken into account. Amounts
of deformations will be different for different images, because of the existing difference
in $R.$

The analysis of focusing action of a GL with an arbitrary mass distribution allows to
argue, that the GL acts as a filter, transmitting low frequencies (slow brightness
variations) with no distortion, and "rejecting" the higher ones, -- rapid variations.

Specific durations of the source brightness variations, $T_s$, for which a smoothing action
of a GL should be taken into accout, can be determined as $T_s<r_g\nu_0/c,$ ($\nu_0
\tau_g<1).$ For example, for a regular model of Q2237+0305, supposing the mass of the
galaxy nucleus $M\sim10^{10}M_{\sun}, (r_g\sim3\cdot10^{10}km)$, and choosing the
characteristic angular size of the source to be equal to $\nu_0\sim10^{-4},$ we get the
estimate for $T_s < 10 s.$ Observations of the Einstein Cross in the IR spectral range,
(Agol, Jones \& Blaes 2000), allow to estimate $\nu_0$ as $10^{-5}<\nu_0<10^{-1}.$ Having
adopted, e.g., $\nu_0\sim10^{-2},$ we get $T_s<10^3 sec.$

\vspace{1cm}
\centerline{\bf Figure captions}
\vspace{0.5cm}
\noindent Fig 1. Mutual locations of a source, $S$, the observer $P$, a macrolens GL, 
and a microlens $m$.\\ 

\noindent Fig 2. Critical curves (1, 2, 3) and caustics ($1\arcmin, 2\arcmin, 3\arcmin$) 
for a microlens location a) outside, b) inside and c) exactly at the critical curve of 
the macrolens (3). The curves (1, 1\arcmin) are for the microlens displacement along the 
X axis, and (2, 2\arcmin) - along the Y axis; 3\arcmin is a macrocaustic.\\

\noindent Fig. 3. Deformations of a macrolens critical curve $\Delta\nu(\varphi)=\nu_{cr}
(\varphi)-\nu^M_{cr}(\varphi)$ for different distances $\vec\nu_m=(\nu_m, \varphi_m=\pi)$ 
between a microlens and the macrolens critical curve.\\ 

\noindent Fig 4. Critical curves (left) and caustics (right) of the discussed macrolens 
model of Q2237+0305, with 300 microlenses, distributed homogeneously within the strip 
of $2\alpha$ width along the macrolens critical curve $\vec\nu^M_{cr}$. \\

\noindent Fig 5. Difference of the critical curves (2) from those calculated in the 
linear approximation (1), for $\Gamma=10^{-9}$, and for the microlens locations a)outside, 
b)inside, and c)exactly at the critical curve of the macrolens (3). The microlens 
coordinates are (0, 0).\\ 

\noindent Fig 6. Behavior of $Q(\tau\arcmin)=\sqrt{2\pi} J_p(\tau\arcmin)/\tilde J_0q$ 
at different values of $R$.\\ 

\noindent Fig 7. Individual (left) and summarized (right) light curves for the values 
of parameter $\tau_g=0.1$.\\

\noindent Fig 8. A total light curve for the value of $\tau_g=600$. 

\vspace{1.0cm}
{Figures 1--8  are available in "gif" format from:}\\
\vspace{1.0cm}
\centerline
{http://arXiv.org/ps/astro-ph/}

\end{document}